\documentclass[10pt,twocolumn,letterpaper]{article}
\usepackage{spconf,amsmath,graphicx}
\usepackage{mdwmath}
\usepackage{mdwtab}
\usepackage{url}
\usepackage{amsfonts}
\usepackage{amssymb}

\title{Leveraging GANs to Improve Continuous Path Keyboard Input Models}

\name{Akash Mehra, Jerome R. Bellegarda, Ojas Bapat, Partha Lal, Xin Wang}

\address{Apple Inc., Cupertino, California 95014, USA}

\begin{document}
\maketitle

\begin{abstract}
  Continuous path keyboard input has higher inherent ambiguity than
  standard tapping, because the path trace may exhibit not only local
  overshoots/undershoots (as in tapping) but also, depending on the
  user, substantial mid-path excursions. Deploying a robust solution
  thus requires a large amount of high-quality training data, which is
  difficult to collect/annotate. In this work, we address this
  challenge by using GANs to augment our training corpus with
  user-realistic synthetic data.  Experiments show that, even though
  GAN-generated data does not capture all the characteristics of real
  user data, it still provides a substantial boost in accuracy at a
  5:1 GAN-to-real ratio. GANs therefore inject more robustness in the
  model through greatly increased word coverage and path diversity.
\end{abstract}


\section{Introduction}
Entering text on a mobile device involves tapping a sequence of
intended keys on a soft keyboard of limited size. Continuous path
input, where users keep sliding their finger across the screen until
the intended word is complete, offers an alternative input modality
\cite{zhai}. After users gain proficiency with such an option, they
often find entering words with one single continuous motion across the
keyboard easier and faster \cite{zhai2}, \cite{zhai3}. Just like for
regular predictive typing, recognition relies on robust pattern
matching enhanced with a statistical language model in order to
predict the intended word \cite{survey}.

\begin{figure}[t]
\vspace{-5truecm}
\hspace{-1.3truecm}\includegraphics[width=1.3\hsize]{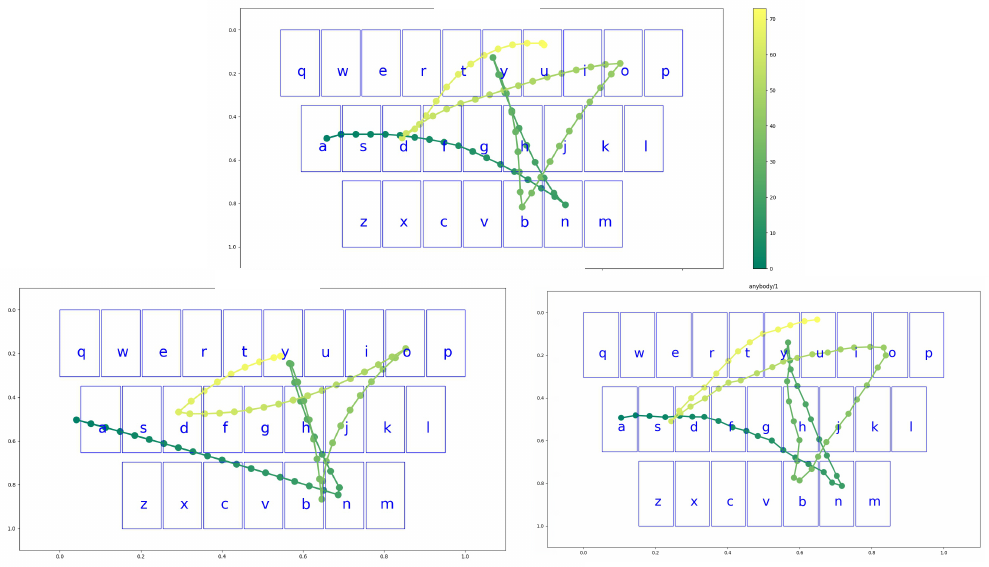}
\vspace{-5.5truecm}
\caption{\sl Path visualization for input word ``{\bf
    anybody}''---color changes from green to yellow with time. Typical
  user path (top), programmatically generated synthetic path (bottom
  left), and GAN-generated synthetic path (bottom right).}
\label{fig1}
\end{figure}

Existing supervised solutions (e.g., \cite{lstm}) based on recurrent
neural networks (RNNs) call for a large annotated corpus of paths,
ideally associated with every token in the supported lexicon. Due to
prohibitive collection and annotation costs, however, the size of that
training corpus is rarely large enough to achieve the required level
of robustness. This observation has prompted a number of
investigations into programmatically generating paths that could be
used as proxies for real user-generated paths. In \cite{lstm}, for
example, the authors generated synthetic paths by connecting the
characters within a word using an algorithm that minimizes jerk
\cite{jerk}, an approach inspired by human motor control theory (cf.,
e.g., \cite{motor}, \cite{motor2}).

Typically, the parameters of the synthesis algorithm are tuned
manually until generated paths look ``similar enough'' to real user
paths (based on human judgments of a small number of paths
\cite{tuning}).  The resulting synthetic paths adequately convey the
associated words, and can even incorporate such artifacts as
undershooting and overshooting some target keys. However, they are
intrinsically restricted in their expressiveness, and do not fully
capture the variability of user paths. To illustrate, Fig.~1 shows a
typical user path (top) and synthetic path (bottom left) for the word
``{\bf anybody}.''

In this paper, we describe a more flexible approach relying on
generative adversarial networks (GANs) \cite{gan}, \cite{gan2}. Given
an initial synthetic path produced with simple cubic splines
\cite{motor}, we transform it in such a way that it conforms to the
kind of user idiosyncrasies observed across the entire set of real
user paths available. This problem can be viewed as an instance of
style transfer, where the goal is to synthesize a given style of
expression while constraining the synthesis to preserve some original
content (cf. \cite{simgan}). The kind of path that results is
illustrated at the bottom right of Fig.~1. GAN generation tends to
more faithfully render human-like artifacts, resulting in better
proxies for real user-generated paths.

The paper is organized as follows. In the next section, we describe
the multi-task bidirectional long short-term memory (bi-LSTM)
architecture we adopted for continuous path style transfer. In
Section~4, we specify the proper objective function to use.  In
Section~5, we discuss our experimental setup, results observed, and
insights gained. Finally, in Section~6 we conclude with a summary and
further perspectives.

\begin{figure}[t]
\centering
\includegraphics[width=0.9\hsize]{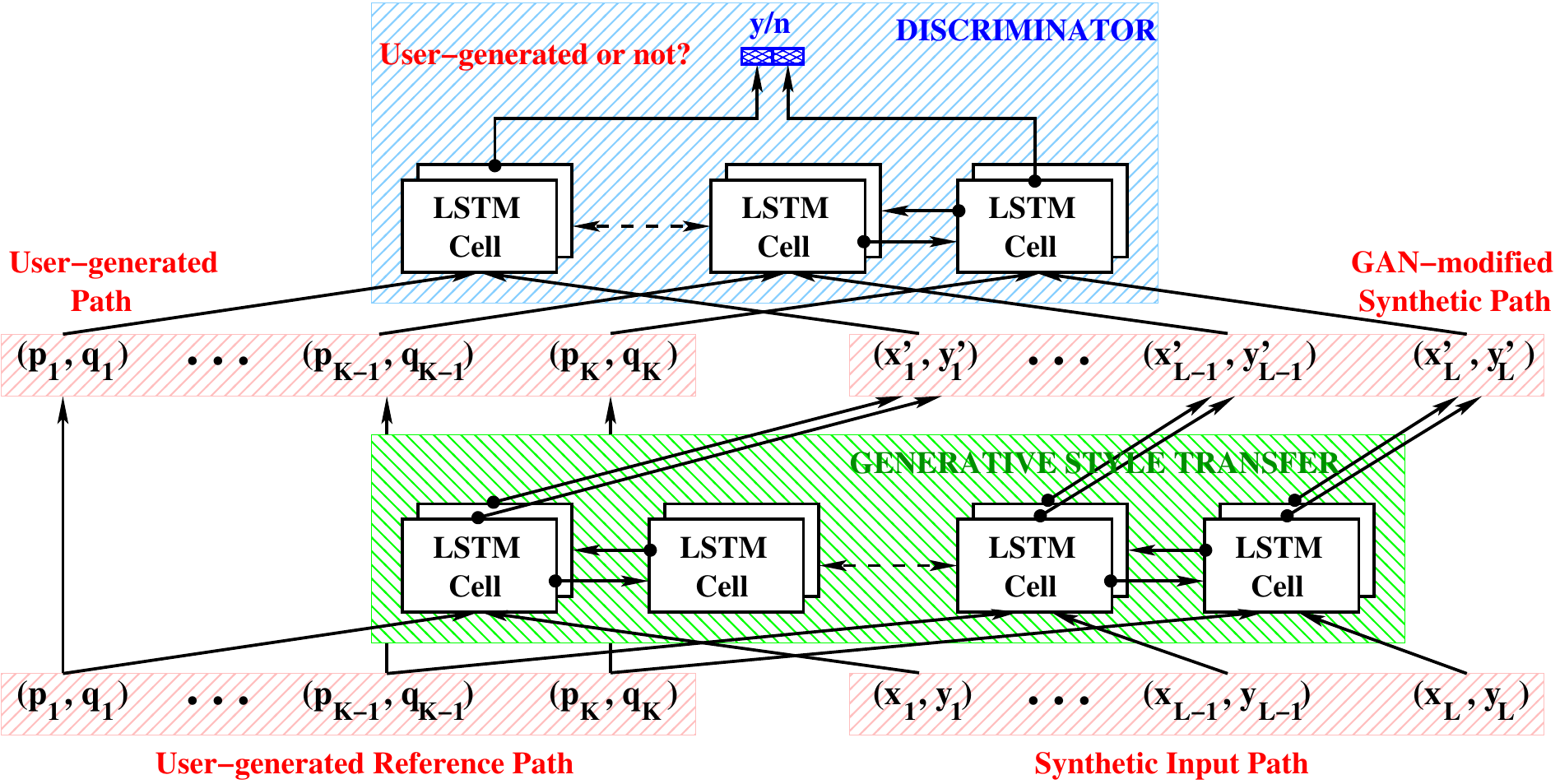}
\vspace{-2truemm}
\caption{\sl GAN architecture for path style transfer.}
\label{fig2}
\end{figure}

\section{Style Transfer Architecture}
Style transfer has been an active area of research in computer vision,
and there is a large body of work on image style transfer: cf., e.g.,
\cite{gatys}--\cite{isola}. The basic idea is to bias the generative
model in GANs according to the desired style of image
\cite{cvpr}. Given the inherent sequential nature of paths, RNNs are
more appropriate for path style transfer than the convolutional neural
networks used in image style transfer. Hence the architecture
illustrated in Fig.~2. Both generative and discriminative models are
realized via bi-LSTMs to avoid vanishing gradients \cite{vanish}. In
each case only one LSTM layer is shown, but in practice it is extended
to a deeper network.

In Fig.~2, an initial synthetic input path $X$ (represented by a
sequence of sampled points $\{(x_1, y_1) \ldots (x_L, y_L)\}$) is
transformed into a ``more human-like'' synthetic path
$Y = \{(x'_1, y'_1 ) \ldots (x'_L, y'_L)\}$ on the basis of an
available set of user-generated reference paths
$P = \{(p_1,q_1) \ldots (p_K, q_K)\}$, which are collectively
representative of a range of observed user idiosyncrasies
(``style''). Both bi-directional LSTMs leverage knowledge of the
entire path under consideration: the generative model to suitably
shift each point so as to make the whole path more congruent to one
generated by a user, and the discriminative model to decide whether
the path is user-generated (as in
$P = \{(p_1, q_1) \ldots (p_K, q_K)\}$) or not (as in
$Y = \{(x'_1, y'_1) \ldots (x'_L, y'_L)\}$). Fig.~2 promotes the
generation of user-realistic paths, because the discriminator will
eventually abstract out user behaviors observed in the $P$ paths and
thus will tend to force the generator to discard transformations that
do not account for such behaviors.

The architecture of Fig.~2 is not completely satisfactory, however,
because there is no guarantee that the transformed path will still be
associated with the correct input word. It could be, for example, that
the original synthetic path becomes virtually indistinguishable from a
real user path corresponding to a different word, which would lead to
a loss of discriminability between words.  This realization caused us
to add a path recognition module, resulting in the architecture
depicted in Fig.~3. By taking into account the built-in constraints
enforced by the classifier, this network is more likely to abstract
out, from the broad inventory of paths occurring in the reference
corpus, those discriminative elements of user generated paths that are
most relevant to the current word.

\begin{figure}[t]
\centering
\vspace{-1truemm}
\includegraphics[width=1.01\hsize]{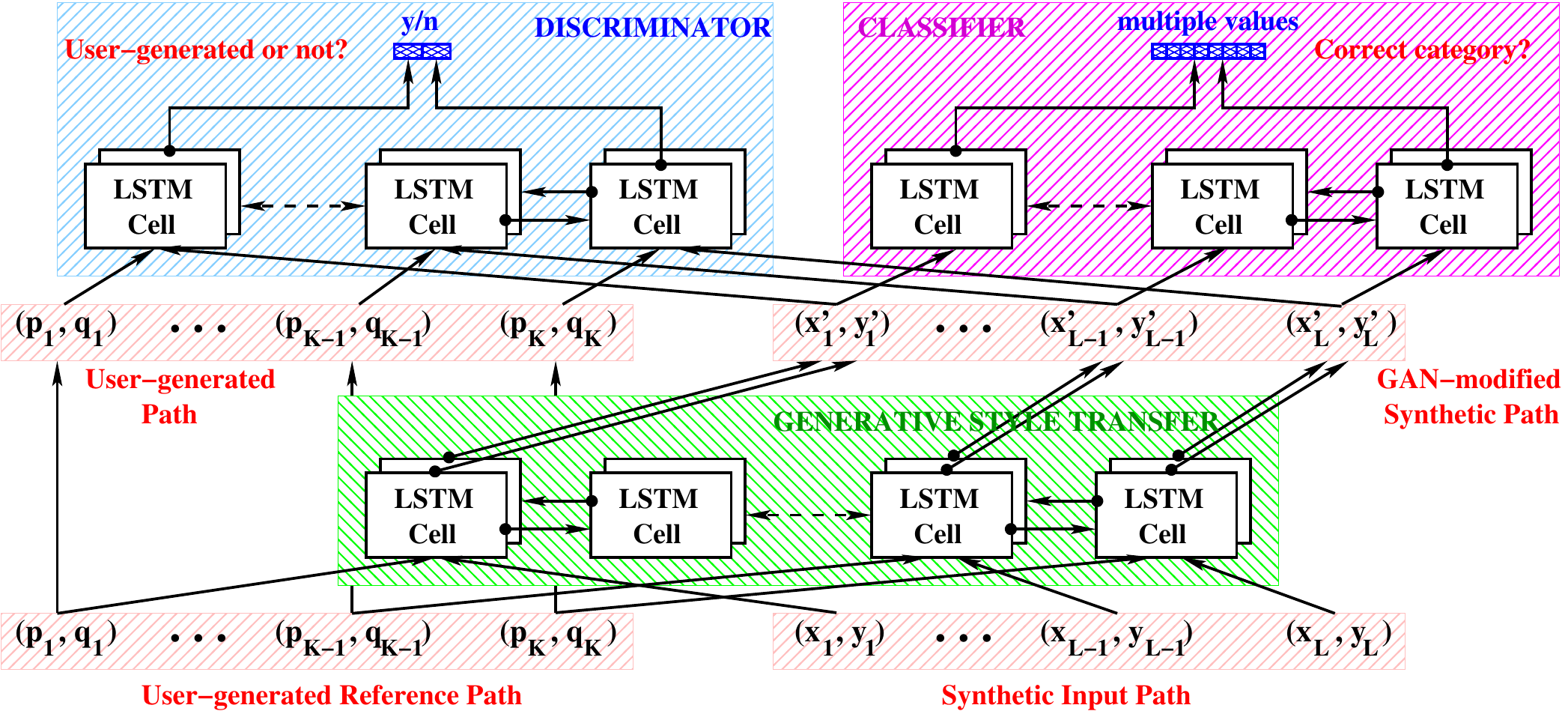}
\vspace{-0.6truecm}
\caption{\sl Multi-task GAN architecture for path generation with
  joint style transfer and classification.}
\label{fig3}
\end{figure}

Compared to Fig.~2, every transformed path $Y$ generated by the style
transfer model is passed not just to the discriminator, but also to a
classifier which verifies that the decoded category is indeed still
associated with the current word. That way, we entice the generative
model to produce a path which is not just similar to a real user path,
but also generally congruent with typical paths for the current word
as observed in the entire reference corpus.  The rest of the system is
as in Fig.~2. In particular, the discriminative model still receives
an input consisting of either a user-generated path $P$ or a synthetic
path $Y$, which implies a multi-task objective function.

\section{Multi-Task Objective Function}
Generative adversarial transfer operates under the assumption that any
path $P$ from the reference corpus is drawn from a true distribution
$\mathbb{D}$ representative of all user-generated behaviors observed
across the available corpus, and that the transformed path $Y$
generated from the initial input path $X$ by the style transfer model
follows a distribution $\mathbb{D'}$ that is ``close'' to $\mathbb{D}$
according to some suitable closeness metric. The objective of GAN
training is therefore for $\mathbb{D'}$ to converge to $\mathbb{D}$.

Looking back at Fig.~2, the generative model $\cal G$ transforms the
input $X$ into $Y= {\cal G}(X)$, taking into account information
provided by $P$ so that $Y$ conforms to the style of $P$. The
discriminator $\cal D$ then estimates the probability that a given
path is drawn from $\mathbb{D}$ rather than $\mathbb{D'}$. Ideally,
${\cal D}(P) = 1$ when $P\sim \mathbb{D}$ and ${\cal D}(Y) = 0$ when
$Y \sim \mathbb{D'}$, with the additional constraint that $Y \approx X$.

This corresponds to a minimax two-player game, in which the generative
and discriminative models $\cal G$ and $\cal D$ are trained jointly
via solving:
\begin{eqnarray}
 \displaystyle \min_{\cal G} \max_{\cal D} \, {\cal K} ({\cal D}, {\cal G}) 
     &=& \mathbb{E}_{P\sim \mathbb{D}} \bigg\{ \log\Big[{\cal D}(P)\Big] \bigg\} \nonumber\\
     &+& \mathbb{E}_{{\cal G}(X)\sim \mathbb{D'}} \bigg\{ \log\Big[1 - {\cal D}\Big({\cal G}(X)\Big)\Big] \bigg\} \nonumber\\
     &+& \mathbb{E}_{{\cal G}(X)\sim \mathbb{D'}} \bigg\{ \Delta\Big[X,{\cal G}(X)\Big] \bigg\} \,,\label{gan2}
\end{eqnarray}
where ${\cal K} ({\cal D}, {\cal G})$ denotes the overall cost
function, and $\Delta[X,Y]$ is a suitable distance metric which is 0
when $X=Y$, and increases away from 0 as the paths $X$ and $Y$ become
more and more dissimilar.  Maximizing (\ref{gan2}) over ${\cal D}$
while minimizing it over ${\cal G}$ ensures that ${\cal G}$ generates
paths that are as maximally similar to $X$ as possible, while looking
like they might have been drawn from the true distribution
$\mathbb{D}$ of user-generated paths. Assuming that ${\cal D}$ and
$\cal G$ comprise a sufficient number of parameters, after enough
training iterations, the distribution $\mathbb{D'}$ will converge to
$\mathbb{D}$ \cite{gan}.  In other words, the network $\cal G$ learns
to synthesize a path $Y={\cal G}(X)$ that eventually looks like it was
user-generated, but still preserves the main characteristics of the
initial path $X$.

The next step is to properly inject the classifier from
Fig.~3. Recognizing the word $w = w_j$ associated with the transformed
path $Y$ involves mapping a potentially long sequence of feature
vectors to a much shorter sequence of characters. A suitable loss
function for this type of sequence classification task is the
Connectionist Temporal Classification (CTC) loss \cite{ctc}. The CTC
loss function trains RNNs on unaligned targets through maximizing the
sum of probabilities of all step-wise sequences that correspond to the
target sequence. Concretely, for a given classifier $\cal C$, this
loss is given as follows:
\begin{equation}
   {\cal L}({\cal C}) \,=\, - \,\log \Big(\sum_{\pi\in{\cal A}(w)}
   \prod_{k=1}^K \,\, o_k^{(\pi)} \Big) \,, \label{ctc}
\end{equation}
where $w$ is the target transcription (word), ${\cal A}(w)$ is the set
of all CTC transcriptions of a target transcription (e.g., for the word
``{\sl data}'', allowable transcriptions may include ``{\sl daata}'', ``{\sl
  datta}'', ``{\sl dddata}'', etc.), and $o_k^{(\pi)}$ denotes the
output of the LSTM at time step $k$ for a particular CTC transcription
$\pi$ of the target transcription of length $K$ characters.

Computing the CTC loss (\ref{ctc}) typically involves inserting blanks at the
beginning, between symbols, and at the end of the target
transcription. The forward-backward algorithm can then be used to
extract all possible transcriptions $\pi$. The desired output follows
after removing blanks and repetitions. Using the CTC loss function
during training thus makes it possible to train the network to output
characters directly, without the need for an explicit alignment
between input and output sequences.

The final step is to combine the two objective functions above to make
sure that the optimal generated path is indeed still associated with
the current word. In practice, to train the network of Fig.~3, we
therefore consider a linear interpolation of the two objective
functions:
\begin{equation}
   {\cal M}({\cal C}, {\cal D}, {\cal G}) \,=\, \lambda \cdot {\cal K} ({\cal D}, {\cal G}) \,+\,
                     (1-\lambda) \cdot {\cal L}({\cal C}) \,, \label{combined}
\end{equation}
where the scalar interpolation coefficient $\lambda$ is a tunable
weighting parameter adjusting the contribution from the main (GAN)
and auxiliary (classification) tasks.

After such multi-task adversarial training is complete, the
discriminative model has learned to abstract out user idiosyncrasies
observed in the reference corpus, so the generated path $Y$ ends up
taking into account the desired range of user behaviors, while still
preserving the main characteristics of the input content $X$ (and
ideally behaving in the same way regarding recognition
accuracy). Thus, the generative network in Fig.~3 in principle leads
to the most realistic rendering of input paths given the reference
corpus.

\section{Experimental Results}
We conducted continuous path recognition experiments using models
trained on a variety of corpora. We drew from a set of 2.2M user paths
(referred below with the prefix ``{\bf U}'') covering 55K English
words collected from a diversity of users in a variety of
conditions. Specifically, 665 users (roughly half males, half females)
ranging in age from 18 to 70 years produced paths on 6 layouts with
various screen sizes. Thus each participant generated 3300 paths on
the average. Approximately half of the paths were generated using the
thumb finger, with the other half generated with the index finger. In
line with \cite{azenkot}, the participants were chosen to attain a
proportion of left-handed users of about 20\%.

We then generated a comparable set of initial synthetic paths obtained
via cubic splines \cite{motor}, referred below with the prefix ``{\bf
  S}.''  Finally, we also used these synthetic paths as initial
exemplars for style transfer, using the multi-task GAN style transfer
architecture of Fig.~3, where the GAN was trained using 1.1M user
paths. This led to the generation of a comparable set of more
sophisticated GAN-generated paths, referred below with the prefix
``{\bf G}.'' For test data, we collected a corpus of
59,469 user paths covering approximately 25K English words included
in the 55K inventory above. To better isolate modeling performance
across different training compositions, we measured Top-1 recognition
accuracy, and thus deliberately ignored language model rescoring.

The results of our experiments are summarized in
Table~\ref{table1}. As baseline, we trained models on 1.1M paths from
each set (top 3 rows, labelled {\bf U1}, {\bf S1}, and {\bf G1}). Then
we analyzed what happens when doubling (next 3 rows, labelled {\bf
  U2}, {\bf U1+S1}, and {\bf U1+G1}) and tripling (following 3 rows,
labelled {\bf U1+S1+G1}, {\bf U1+G2}, and {\bf U2+G1}) the amount of
training data, via folding in either more user paths or more
synthetic/GAN paths. Finally, we investigated whether the same trends
hold when training on 4.4M paths, only half of which are user paths
(following two rows, labeled {\bf U2+S2} and {\bf U2+G2}) and when
training on approximately 13M paths, only about 1/6 of which are user
paths (last row, labeled {\bf U2+G10}).

\begin{table}[htb]
\caption{\sl Continuous path recognition results using a variety of
  training compositions, given an underlying lexicon of 55K English
  words. The test corpus comprises 59,469 paths covering 24,002
  words. No language model is used.}
\vspace{2truemm}
\renewcommand{\arraystretch}{1.3}
\centering
\begin{tabular}{|l|c|}
\hline
Training Composition (Number of Paths)   & Top-1 Acc.\\
\hline
{\bf U1} (1.1M user only)                                    & {\bf 58.5\%}   \\
{\bf S1} (1.1M synt. only)                                    & 35.0\%   \\
{\bf G1} (1.1M GAN only)                                    & 33.7\%   \\
\hline
{\bf U2} (2.2M user only)                                     & 62.2\%   \\
{\bf U1+S1} (1.1M user + 1.1M synt.)                  & 57.6\%   \\
{\bf U1+G1} (1.1M user + 1.1M GAN)                  & {\bf 62.4\%} \\
\hline
{\bf U1+S1+G1} (1.1M each of user, synt., GAN)  & 61.4\%   \\
{\bf U1+G2} (1.1M user + 2.2M GAN)                   & 63.5\%   \\
{\bf U2+G1} (2.2M user + 1.1M GAN)                   & {\bf 64.5\%}   \\
\hline
{\bf U2+S2} (2.2M user + 2.2M synt.)                   & 59.5\%   \\
{\bf U2+G2} (2.2M user + 2.2M GAN)                    & {\bf 65.8\%}   \\
\hline 
{\bf U2+G10} (2.2M user + 10.8M GAN)                & {\bf 66.8\%}   \\
\hline 
\end{tabular}
\vspace{-2truemm}
\label{table1}
\end{table}

The most salient observation is that adding GAN-generated data is much
more effective than adding less user-realistic synthetic
data. Interestingly, GAN-generated paths by themselves prove no more
effective than synthetic paths by themselves (compare {\bf U1}
vs. {\bf S1} vs. {\bf G1} in the top 3 rows). This may be traced to a
lack of diversity in data generation, or possibly a failure to capture
all relevant user artifacts. However, when used as a complement to
user data (compare {\bf U2} vs. {\bf U1+S1} vs. {\bf U1+G1} in the
next 3 rows), GAN data clearly outperforms synthetic
data---essentially providing the equivalent of training on user data
only in terms of accuracy.

This trend is confirmed when folding in more data (compare {\bf
  U1+S1+G1} vs. {\bf U1+G2} vs. {\bf U2+G1} in the next 3 rows):
synthetic data completely fails to help over the (1.1M+1.1M) user/GAN
training scenario, while folding in either more user or more GAN data
substantially improves the outcome. Note that user data seems to
retain its expected edge, as the (2.2M+1.1M) configuration still
outperforms the (1.1M+2.2M) configuration by 1\% absolute. Again, this
may be due to the inherent difficulty of capturing rarely observed
user artifacts using a GAN framework. Finally, even better results are
obtained in the (2.2M+2.2M) user/GAN configuration, which provides a
3.6\% absolute accuracy boost over the baseline 2.2M user training,
and the best results are obtained when folding even more GAN data: in
the (2.2M+10.8M) user/GAN configuration, we reach a 4.6\% absolute
boost in accuracy, showing that it is possible to exploit to good
effect many multiples of the user-generated data available.

\begin{figure}[t]
\centering
\vspace{-1truemm}
\includegraphics[width=0.9\hsize]{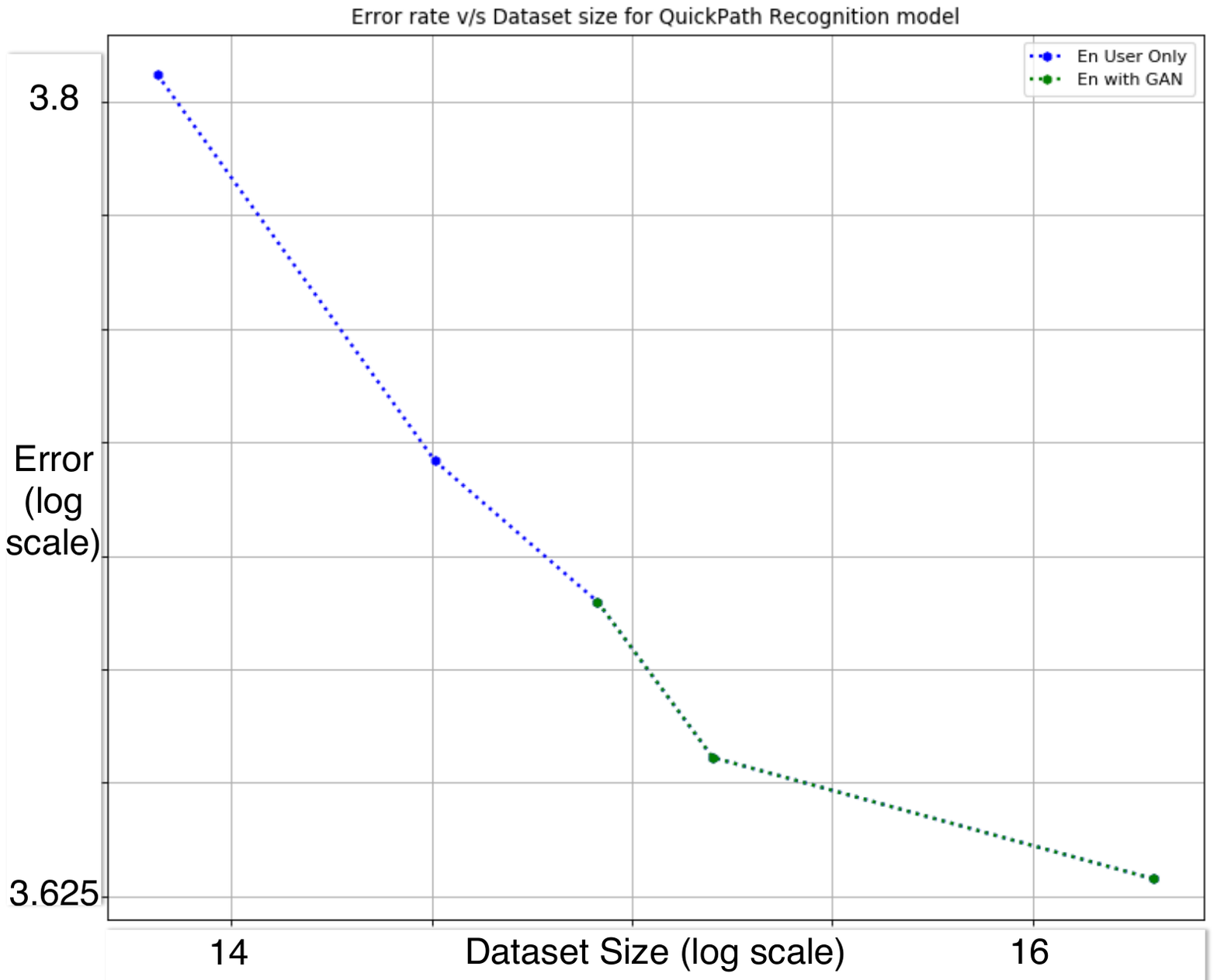}
\vspace{-0.4truecm}
\caption{\sl Log-log plot of continuous path recognition learning
  curve derived from Table~1. Blue: user data only, green: GAN augmentation.}
\label{fig4}
\vspace{-0.2truecm}
\end{figure}

Fig.~4 displays the associated learning curve, i.e., how the error
scales with data size. The ``steepness'' of that learning curve (on a
log-log plot) conveys how quickly a model can learn from adding more
training samples \cite{baidu}. For a given model architecture, it also
implicitly reflects the quality of the added samples. Given than the
green (GAN) slope is slightly less steep than the blue (user) slope,
it appears that adding GAN-generated data is somewhat less effective
than adding real user data. In other words, GAN-generated data only
captures {\it some} of the desired user characteristics.  Still, the
learning curve of Fig.~4 bodes well for reducing data collection costs
when extending to other languages and scripts.

\section{Conclusion}
In this paper, we have leveraged GANs to synthesize user-realistic
training data for learning continuous path recognition models. This
approach obviates the need to learn the parameters of a dedicated
human motor control model, or to assess the similarity between a
synthetic and user-generated path. Such assessment becomes an emergent
property of the modeling, which in turn enables practical deployment
at scale.

The proposed approach relies on a multi-task adversarial architecture
designed to simultaneously carry out two different sub-tasks: (i)
generate a synthetic path that generally conforms to user
idiosyncrasies observed across reference user-generated paths; and
(ii) verify that recognition accuracy for this synthetic path is not
negatively impacted in the process. After multi-task adversarial
training is complete, the GAN-generated paths reflect a range of
realistic user behaviors while still being aligned with the target
word.

The generated paths are then folded into the continuous path training
corpus, which advantageously increases both word coverage and path
diversity. That way, we emulate the acquisition of many more paths
from many more users and thereby support training of a more robust
model. This solution considerably reduces the collection and
annotation costs typically associated with the large corpus of paths
required (cf., e.g., \cite{lstm}). A further improvement will be to
streamline the two sequential processes involved: path generation in
order to augment the training corpus, followed by standard training of
the path recognition model. In principle, the fact that the path
generation process already comprises a recognition module could enable
consolidation into a unified training performed in end-to-end
fashion.

\end{document}